\documentclass[manuscript,prb,aps]{revtex4}
\usepackage{graphicx}
\usepackage{bm}
\usepackage{color}
\begin{document}
\title{Influence of Dielectric Environment on Role of Spin-Orbit Interaction for Image Potentials}
\author{Godfrey Gumbs$^1$,  Oleksiy Roslyak$^2$, Danhong Huang$^3$, Antonios Balassis$^4$}
\address{$^1$Department of Physics and Astronomy, Hunter College of the
City University of New York, 695 Park Avenue, New York, NY 10065, USA\\
$^2$Center of Integrated Nanotechnology, CINT, Los Alamos, NM, 87545, USA\\
$^3$Air Force Research Laboratory, Space Vehicles Directorate, Kirtland Air Force Base, NM 87117, USA\\
$^4$Physics Department, Fordham University, 441 East Fordham Road, Bronx, NY 10458, USA}
\date{\today}

\begin{abstract}
We present a formalism for calculating the image
potential for a two-dimensional electron gas (2DEG)
with Rashba spin-orbit interaction (SOI) as well as for a
2D topological insulator (TI).
The formalism is further generalized for including the Coulomb coupled multiple layers.
Roles of broken inversion symmetry near the surface and
the dielectric environment are investigated by using a surface-response function.
The insignificant role of SOI in 2DEG is dramatically enhanced in TI by selecting a small relative permittivity $\epsilon_b$
for the dielectric environment.
Manipulating $\epsilon_b$ is proven to provide an
efficient way to drive electrons with opposite spins into two different
integral quantum Hall states.
The prediction made in this paper is expected to be experimentally observable for a
2DTI system, such as Bi$_2$Se$_3$, with a helical spin behavior and
a dominant linear Rashba SOI-like term in the energy dispersion.
\end{abstract}

\_\hrulefill\_\vspace {0.18in}


\maketitle

\section{Introduction}
\label{sec1}

In recent times, there has been a considerable amount of research
studies on the role played by the Rashba\,\cite{rashba} spin-orbit
interaction (SOI) on the collective-excitation properties of the
two-dimensional electron gas (2DEG) formed in a semiconductor
heterostructure. Special focus was placed on its affluence of plasma excitations\,\cite{gg,manvir1,manvir2,wenxu,1,2,3,exact1,exact2} and
electron transport\,\cite{3}. It has been predicted theoretically
that plasmon  propagation becomes tunable in the presence of an
external electric field via SOI. This provides a way for transmitting
quantum information in a quantum device and opens up a new possibility
for the so-called plasmon-field effect transistor. Experimentalists are
specifically interested  in exploiting the Rashba  SOI in InAs or
InGaAs so as to gain control over the spin population by increasing a
lateral or gate field and then pumping electrons. There were some papers\,\cite{M1,M2,M3,M4,M5} on ballistic quantization in InAs, even
though the mobility was much less than GaAs. However, in a theoreti
cal study, we found that SOI plays a significant role in the conductance
quantization of quantum wires\,\cite{dh1,dh2}.
\medskip

In Refs.\,[\onlinecite{exact1,exact2}],
the authors evaluated charge and spin density response functions\,\cite{gg}
of a 2DEG with Rashba spin-orbit coupling at finite momenta and frequencies.
The polarization function is the basis for calculating the plasmon dispersion
and the quasi-classical approximations.
Additionally, one can employ the polarization function to obtain the
surface-response function which is a crucial ingredient in the analysis
of the electronic properties of single and Coulomb coupled 2DEG
layers\,\cite{book}.
\medskip

Here, we investigated the effect of
the Rashba SOI on the image potential for a single-layer 2DEG
as well as for two layers which are Coulomb coupled.
Using the static limit for the surface-response function with
and without SOI calculated self-consistently, we were able to analyze
the role played by adjoining dielectric media on the image potential.
In 2DEG, the linear-$k_{\|}$ (in-plane wave vector) term (hereafter
referred as Rashba term) lifts the spin degeneracy of the dispersion
away from $k_{\|}=0$. But it only serves as a small perturbation to
the dominant quadratic mass term $\sim k^2_{\|}$. In 2D topological
insulators (TIs)\,\cite{Science},  such as Bi$_2$Se$_3$, their roles are inverted,
that is, the Rashba term dominates the dispersion providing helical flavor to the
topological conductance and valence surface states.
This type of  helical behavior also occurs in
conventional 2DEG with SOI due to  broken inversion symmetry
near the sample surface.  We compare the image potentials of both
2DEG and TI when the dielectric environment is varied.
Related image-potentials for the semi-infinite metal/vacuum
interface as well as for graphene have been studied\,\cite{17,image1,image2}.
Recently, the image states have been observed in pyrolytic
graphite\,\cite{18} and metal supported  graphene\,\cite{19}.
We analyze how the SOI and the strong interlayer Coulomb coupling
affect the image states.
The  image states for the double layer are formed through the interlayer
hybridization of the  image-potential state in individual 2DEG layer.
The situation may be compared with that of double-layer graphene\,\cite{4+}.
\medskip

If a stationary external change, i.e. having  velocity ${\bf v}=0$,
is introduced to our two-dimensional electronic systems,
we need only consider the static limit with
$\omega\sim{\bf q}\cdot{\bf v}\rightarrow 0$, where
${\bf q}$ is the wave vector of an induced collective excitation of
the electron gas. Therefore, we can neglect the force contribution from
the change of a vector potential with time and simply assume a
zero Lorentz force for the external charge. As indicated previously\,\cite{mop},
there exists a magnetoelectric polarizability associated with
3D topological insulators due to the so-called axion electrodynamics.
As a result of this, the displacement field will also depends on the
magnetic field, and the magnetic field in turn depends on the electric
field as well. However,  Poisson's equation for the scalar potential
still holds if we select the Coulomb gauge for our system. Additionally,
the equation for the vector potential will maintain its form
if the material considered is non-magnetic. In this sense, the
image potential for the  two-dimensional electron gas and topological
insulator models discussed in Sec.\,\ref{sec2} is well defined.
\medskip

In addition, the image states of metallic carbon nanotubes\,\cite{7}
and double-wall non-metallic nanotubes\,\cite{8,8b} were investigated.
Experimental work\,\cite{8c} includes photoionization\,\cite{8c+}
and time-resolved photoimaging of image-potential states in carbon
nanotubes\,\cite{8cc+}. There  has been general interest\,\cite{8c++}
in these structures because of electronic control on the nanoscale
using image states. This has led to wide-ranging potential applications
including field ionization of cold atoms near carbon nanotubes\,\cite{8d},
and chemisorption of fluorine atoms on the surface of carbon
nanotubes\,\cite{8e}. We anticipate that the image states we investigate
would lead to the experimental study of spin-orbit effects on
electronic control devices, for example. Furthermore, the
significance of the role played by the image-potential  should
become pronounced under suitable conditions which are discussed in
the present paper.
\medskip

In the rest of the paper, we first derive a formalism in Sec.\,\ref{sec2} for calculating the image potential by a point charge in terms of a surface-response function.
The surface-response function is further related to a single-particle density-density response function for both single- and double-layer systems.
In Sec.\,\ref{sec3}, the density-density response function is calculated explicitly for both 2DEG and TI systems.
The numerical results for image potentials are compared in Sec.\,\ref{sec4} for both single- and double-layer 2DEG and TI as well embedded in different
dielectric environments and with or without a Rashba SOI term. Finally, a brief conclusion is drawn in Sec.\,\ref{sec5} with a remark.

\section{The image potential formalism}
\label{sec2}

Consider a stationary external point charge  $-e$ located at $\textbf{r}_0=(0,0,z_0)$ with $z_0>0$
on the polar $z$ axis in vacuum. We assume that there is a material  with
background dielectric constant $\epsilon_b$  in the region $z<0$.
The  external potential at  $\textbf{r}$ due to the presence of such a
point charge can be obtained by solving Poisson's equation

\begin{equation}
\nabla^2\phi_{\rm ext}(\textbf{r})=-\frac{e}{\epsilon_0}\,\delta(\textbf{r}-\textbf{r}_0)\ ,
\label{poisson}
\end{equation}
where the standard units is adopted with $\epsilon_0$ denoting the permittivity
of free space. For $z<z_0$,  we obtain from Eq.\,(\ref{poisson})

\begin{equation}
\phi_{\rm ext}(\textbf{r})= -\frac{e}{4\pi\epsilon_0}
\int d^2\textbf{q}_{\|} \ e^{-q_{\|}(z_0-z)}
\left( \frac{1}{2\pi q_{\|}}  \right) e^{i{\bf q}_{\|}\cdot{\bf r}_{\|}}\ .
\end{equation}
Also, for $z\geq 0$, the induced potential  is given by

\begin{equation}
\phi _{\rm ind}(
\textbf{r})=\frac{e}{4\pi\epsilon_0}\int d^2\textbf{q}_{\|} \ e^{-q_{\|}(z_0+z)}
\left( \frac{1}{2\pi q_{\|}}\right)\,g(q_{\|},\,\omega=0)\,e^{i{\bf q}_{\|}\cdot{\bf r}_{\|}}\ ,
\end{equation}
where $g(q_{\|},\,\omega=0)$ is the static surface-response function to be determined.
Therefore, the force exerted on the external charge due to the induced charge in
the medium is given by

\begin{eqnarray}
\textbf{F}_{\rm ind}&= &\left.e\,\hat{\bf z}\,\frac{\partial}{\partial z}\phi_{\rm ind}(\textbf{r})
\right|_{z=z_0,\,{\bf r}_{\|}=0}
\nonumber\\
&=&  - \frac{e^2}{4\pi\epsilon_0}
\int \frac{d^2\textbf{q}_{\|}}{2\pi} \ e^{-2q_{\|}z_0}
\,g(q_{\|},\,\omega=0)\,\hat{\bf z}
\equiv\frac{\partial}{\partial z_0}{\cal U}_{\rm im}(z_0)\,\hat{\bf z}\ .
\label{relation}
\end{eqnarray}
The relation in Eq.\,(\ref{relation}) defines the image potential as

\begin{eqnarray}
{\cal U}_{\rm im}(z_0)&=&\frac{1}{2}
\left(  \frac{e^2}{4\pi\epsilon_0}\right)
\int_0^\infty dq_{\|}\
e^{-2q_{\|}z_0}
\,g(q_{\|},\,\omega=0)
\nonumber\\
&=&\left(  \frac{e^2}{4\pi\epsilon_0}\right)
\frac{1}{4z_0} -\left(  \frac{e^2}{4\pi\epsilon_0}\right)
\frac{1}{2} \int_0^\infty dq_{\|} \
e^{-2q_{\|}z_0}\,\left[1-g(q_{\|},\,\omega=0)  \right]\ .
\label{U-image}
\end{eqnarray}

If we assume that there is a layer of 2DEG at $z=0$ and a second
layer at distance $d$ from the first with material having dielectric
constant $\epsilon_b$ between but vacuum for $z>0$ and $z<-d$, then
the dynamical surface-response function for this double layer is given by

\begin{equation}
g_{\rm im,DL}(q_{\|},\,\omega)=1+2\,\frac{1+\epsilon_b-
\alpha_{\rm L}/(4\pi\epsilon_0\,q_{\|})
-[1-\epsilon_b-\alpha_{\rm L}/(4\pi\epsilon_0\,q_{\|})]e^{-2q_{\|}d}}
{[1-\epsilon_b-\alpha_{\rm L}/(4\pi\epsilon_0\,q_{\|})]^2e^{-2q_{\|}d}
-[1+\epsilon_b-\alpha_{\rm L}/(4\pi\epsilon_0\,q_{\|})]^2}\ ,
\label{eels31}
\end{equation}
where $\alpha_{\rm L}\equiv(4\pi e^2/\hbar^2)\,\Pi^0(q_{\|},\,\omega)$
and $\Pi^0(q_{\|},\,\omega)$ is the single-particle
density-density response function.
\medskip

In the limit of $q_{\|}d\to\infty$, we obtain the surface response from
Eq.\,(\ref{eels31}) for a single layer as

\begin{equation}
g_{\rm im,SL}(q_{\|},\,\omega)=1-\frac{2}
{1+\epsilon_b-\alpha_{\rm L}/(4\pi\epsilon_0\,q_{\|})}\ , \label{eels31c.1}
\end{equation}
which was previously derived in the paper by Persson\,\cite{bnj}.
We find that if we replace the vacuum regions in our calculations above
by a material with background dielectric constant $\epsilon_1$ and
the material between the double layers by a material with
background dielectric constant $\epsilon_2$, then the corresponding
surface-response function would still be given by Eq.\,(\ref{eels31}),
but with $\epsilon_b$ defined as the ratio of the two dielectric constants,
i.e., $\epsilon_b\equiv \epsilon_2/\epsilon_1$. This then allows us
to vary the nature of the dielectric environment to study the
corresponding effect on the image potential.
\medskip

To proceed further with our calculations, we need to calculate the
polarization function. A natural first step in this direction is
the energy eigenstates which we turn to in Sec.\,\ref{sec3}. However,
if we neglect the effect from the density-density response on the
surface-response function (i.e. by taking $\alpha_{\rm L}=0$), the image potential
(in units of $-e^2/(2\epsilon_0)\,\sqrt{2 \pi n_e
+(\Delta_{\rm R} m^\ast/\hbar^2)^2}$ with Rashba parameter $\Delta_{\rm R}$ and electron density $n_e$)
for single and double layer configurations
can be simply calculated as

\begin{eqnarray}
\label{EQ:IM1}
&&\mathcal{U}_{\rm im,SL} \left({z_0}\right) =
\frac{\epsilon_b-1}{2 z_0 \left({\epsilon_b+1}\right)}
\nonumber\\
&& \mathcal{U}_{\rm im,DL} \left({z_0,d}\right)
=\frac{\left({\epsilon_b -1 }\right)}{2 z_0 \left({\epsilon_b+1}\right)^2 \left({d+z_0}\right)}\ ,
\nonumber\\
&\times& \left\{z_0 \left(1- 3
\epsilon_b\right) \, _2 F_1
\left[1, \frac{d+z_0}{d},
\frac{z_0}{d}+2,\left(\frac{\epsilon_b-1}{\epsilon_b+1}\right)^2 \right]
\right.
\nonumber\\
&& \left. -\left(\epsilon_b+1\right) \left(d+z_0\right)
\, _2 F_1
\left[1, \frac{z_0}{d},
\frac{z_0+d}{d},\left(\frac{\epsilon_b-1}{\epsilon_b+1}\right)^2 \right]
 \right\}\ ,
\end{eqnarray}
where $_2 F_1$ is a hypergeometric function and $z_0$ is in units of $1/\sqrt{2 \pi n_e
+(\Delta_{\rm R} m^\ast/\hbar^2)^2}$.

\section{The Polarization Function}
\label{sec3}

For a noninteracting ideal 2D electron or heavy hole (HH) gas
in the $xy$-plane with SO coupling, the single-particle
Hamiltonian is well known\,\cite{rashba}. In fact,
for a given spin, the Hamiltonian leads to a momentum-dependent
force on the electron/hole, somewhat like a
magnetic field. Moreover, the spin-dependence means that the time-reversal
symmetry of SO coupling  is different from a real
magnetic field. For this case we can write down the Hamiltonian as

\begin{equation}
{\cal H}^{(\nu)}= \left[\matrix{-\hbar^2/(2m^\ast)\,\nabla_{\|}^2 &
\Delta_{\rm R}^{(\nu)}\nabla_-\cr
\Delta_{\rm R}^{(\nu)}\nabla_+ &
-\hbar^2/(2m^\ast)\,\nabla_{\|}^2\cr}
\right]\ ,
\label{Hamiltonian-SOI}
\end{equation}
where $m^\ast$ is the effective mass of an electron or HH,
$\nabla_{\|}^2=\partial^2/\partial x^2+
\partial^2/\partial y^2$, $\mp\nabla_\pm=
\partial/\partial x\pm i\,
\partial/\partial y$. In addition, $\Delta_{\rm R}^{(\nu)}$
is the Rashba parameter for an electron or HH system with $\nu=e$ for electrons and $\nu=h$ for HHs.
The energy eigenvalues for the Hamiltonian ${\cal H}^{(\nu)}$ in Eq.\,(\ref{Hamiltonian-SOI})
are, in terms of an in-plane wave vector ${\bf k}_{\|}=(k_x,\,k_y)$,
given by

\begin{equation}
E_{k_{\|},\,\eta}^{(\nu)}=\frac{\hbar^2k_{\|}^2}{2m^\ast}+\eta\,
\Delta_{\rm R}^{(\nu)}\,k_{\|}\ ,
\label{Hamiltonian-SOI-2}
\end{equation}
showing Dirac dispersion and helical spin texture,
with eigenfunctions

\begin{equation}
\psi_{k_{\|},\eta}^{(\nu)}({\bf r})=\left[
\matrix{1\cr
(\eta/k_{\|}^j)\left(k_y-ik_x  \right)^j
\cr}
  \right]
\frac{e^{i{\bf k}_{\|}\cdot {\bf r}}}{\sqrt{A}}\ ,
\label{eigen1}
\end{equation}
where $\eta=\pm$ and ${\cal A}$ is a normalization area. Also, for
a chosen total electron ($n_e$) or hole ($n_h$) density, the spin ``$+$" or
spin ``$-$" carriers will be distributed between the
two sub-bands with density $n_\pm$ determined by

\begin{equation}
\frac{n_\eta}{n_\nu}=\frac{1}{2}-\eta\,C^{(\nu)}\left[\left(1-\frac{n_\eta}{n_\nu}
\right)^{j/2}+ \left(\frac{n_\eta}{n_\nu}
\right)^{j/2}  \right]\ ,
\label{number}
\end{equation}
where $j=1$ for electrons and $j=3$ for HHs, $C^{\rm(e)}=m^\ast\Delta_{\rm R}^{\rm(e)}/(2\hbar^2\sqrt{\pi n_e})$ for electrons and
$C^{\rm(h)}=2m^\ast\Delta_{\rm R}^{\rm(h)}/(\hbar^2\sqrt{\pi n_h})$ for HHs. Both
$C^{(e)}$ and $C^{\rm(h)}$ must be less than $1/2$ to ensure the validity of Eq.\,(\ref{number}).
If $C^{\rm(e,h)}>1/2$, then only the spin-``$-$" sub-band
is occupied.
In such a model, electron spin is conserved, and
there exists a spin current.
An applied electrical field causes oppositely directed
Hall currents with `+' and `$-$' spins, respectively.
The net charge current is zero, while the net spin current
is nonzero, and even becomes quantized. However, in real solids
there is no conserved direction of spin.
Consequently, it is expected that
$\uparrow$/+ and $\downarrow$/$-$
would always mix and the effect arising from the edge will be canceled out.
The model state in this theory is just two copies of the integer quantum Hall
(IQH) state. It was shown  by Kane and Mele\,\cite{Kane-Mele} that, in real
solids with all spins mixed and a zero net spin current,
only some part of physics outcome in this model can survive.
Kane and Mele\,\cite{Kane-Mele} further found a new topological invariant in
time-reversal-invariant systems of fermions.
\medskip

It is known that there exists a surface state, called a ``holographic metal'', in Bi$_2$Se$_3$ family materials,
where the surface state is 2D but still determined by the 3D bulk topological property.
Such surface states can exist only in the range from $k=0$ to $k=k_{\rm c}$ before their merging with (quadratic) bulk states and paring with surface states of the other boundary.
Therefore, the number of Fermi-surface crossing for a single Kramers partner at one surface can be an odd integer, which is different from an even integer in a 1D time-reversal invariant system.

\begin{equation}
{\cal H}_{\rm surf}=\left(C_2+\alpha_2\,M_2\right)k^2\,\mathbf{I}+A_0\,\alpha_1
\left(\boldsymbol\sigma_x k_y-\boldsymbol\sigma_y k_x\right)\ ,
\end{equation}
where the first term is related to quadratic bulk states, $\mathbf{I}$ is the $2\times 2$ unit matrix and
$\boldsymbol\sigma_{x,y}$ are the Pauli matrices. The parameters $\alpha_1,\,\alpha_2$ are phenomenological, taken from experiment and
their values are indicated in the text. The parameters $C_2,\,M_2,\,A_0$
are specific for the material and can be calculated from the four-band
tight-binding model (See Liu, et al., in Ref.\,[\onlinecite{liu2010model}]).
 We also present them in the text for  $Bi_2 Se_3$. The similarity between the surface TI Hamiltonian and that for SOI was indicated in
Ref.\,[\onlinecite{liu2010model}]. The model of Bernevig, Hughes and Zhang\,\cite{BHZ} has been widely used for the TI description.
Of course, one is justified in noticing  the even (SOI) and odd (TI) number
of dispersion crossing points with the Fermi surface. This  arises from the
dominantly quadratic term in the case of the SOI compared with the mostly linear
TI dispersion. That is, for TIs, the quadratic part merges with the bulk valence
band and is subsumed. In the model Hamiltonian, this property is of course not
apparent. However, in the derivation of the polarization  function (See Pletyukhov, et al., in Ref.\,[\onlinecite{exact2}]),
a cut-off function for the wave vector was introduced. The value of the cut-off
is much smaller than the crossing point where the quadratic part creates a non-physical
crossing point. Therefore, there is no error introduced by those possible excitations.
Consequently, in some way, the odd number of crossing points has to be built into the
degree of accuracy for calculating the polarization  function. From a mathematical point of view,
the  crossing points are $k=0$ and $k=2k_{\rm R}$, where the Rashba wave vector is $k_{\rm R} =m^\ast\Delta_{\rm R}$
in Eq.\,(2) of  Ref.\,[\onlinecite{exact2}]. The cut-off in Eq.\,(10) of Ref.\,[\onlinecite{exact2}] is $k_F\mp k_{\rm R}$. The first branch
cut-off $k_F-k_{\rm R}\approx \sqrt{2 m^\ast E_F}$ is much less than $2k_{\rm R}$ so that our
method of calculation is reasonable. But, the second branch which is unphysical for
TI must be removed so that it does not contribute numerically to the static
polarization function.
\medskip

It is interesting to note that  the Hamiltonian describing
surface states for Bi$_2$Se$_3$ TI may also be formally written
in the form given by Eq.\,(\ref{Hamiltonian-SOI}) with
energy eigenvalues in Eq.\,(\ref{Hamiltonian-SOI-2}) for
topological surface states\,\cite{rashba,winkler2003}.
Consequently, we shall refer to the linear term in the
energy dispersion as
a Rashba-like term for TIs.   As a matter of fact,  for the TI with
the material parameters\,\cite{liu2010model}
$\alpha_1= 0.99$, $\alpha_2=-0.15$,
$A_0 = 3.33$\,eV$\cdot$\AA,
$C_2 = 30.4$\,eV$\cdot$\AA$^2$,
$M_2 = 44.5$\,eV$\cdot$\AA$^2$,
we get an effective mass
$m^\ast_{\rm TI}=\hbar^2/[2(C_2+\alpha_2\,M_2)]=0.16058\,m_e$
where $m_e$ is the free electron mass. The Rashba-like parameter
is given by $\Delta_{\rm R}\equiv \alpha_1 A_0$.
The above parameters were confirmed by {\em ab initio\/}
calculations\,\cite{zhang2010first} and experimentally verified
via spin-resolved  photo-emission spectroscopy (SRPES)\,\cite{hsieh2009tunable}.
Note that the sign of $\alpha_1$ is determined by underlying atomic SOI.
\medskip

The interplay between the two quantities, i.e. $m^\ast$ and $\Delta_{\rm R}$,
indicates that the Rashba-like term in TI is dominant
in contrast with conventional 2DEG. As a result, the
surface states show  almost linear dispersion with helical spin
structure, which has the opposite direction for the conduction and
valence bands. Such helical structure is also characteristic for
conventional 2DEG with SOI and arises from broken inversion symmetry
near the sample surface. This symmetry breaking also manifest itself in  the
electron density distribution, as given by Eq.\,(\ref{number}).
\medskip

The linear screening of an external potential $\phi_{\rm ext}({\bf r})$
by the 2DEG (or TI surface states) embedded in a medium with background dielectric
constant $\epsilon_b$ is given by

\begin{equation}
\Phi_{\rm tot}({\bf r},\omega)=\int_{\cal A}  d^3{\bf r}^\prime
\ \epsilon^{-1}({\bf r}, {\bf r}^\prime;\omega)\,\phi_{\rm ext}({\bf r}^\prime)\ ,
\end{equation}
where the inverse dielectric function is expressed in terms of the
density-density response function $<[n({\bf r}^{\prime\prime},t),
n({\bf r}^\prime,0)]_->$ through

\begin{eqnarray}
\epsilon^{-1}({\bf r},{\bf r}^\prime;t)&=&\delta({\bf r}-{\bf r}^\prime)\,\delta(t)
+\frac{1}{i\hbar}\int_{\cal A} d^3{\bf r}^{\prime\prime}\
\frac{e^2}{4\pi\epsilon_0\epsilon_b|{\bf r}-{\bf r}^{\prime\prime}|}
\nonumber\\
&\times& <[n({\bf r}^{\prime\prime},t),
n({\bf r}^\prime,0)]_-> \ .
\end{eqnarray}

In  the self-consistent-field theory, the dielectric function takes in the form of $\epsilon(q_{\|},\,\omega)=1-\alpha_{\rm L}(q_{\|},\,\omega)/(4 \pi \epsilon_0 \epsilon_bq_{\|})$, where the
dynamic polarization function is given by

\begin{eqnarray}
&&\alpha_{\rm L}(q_{\|},\,\omega)= -\ \frac{2\pi e^2}{\cal A}
\sum_{{\bf k}_{\|}}\,\sum_{\sigma,\sigma^\prime=\pm1}\,
\frac{f_0(E^{\rm(e)}_{k_{\|},\sigma})-
f_0(E^{\rm(e)}_{|{\bf k}_{\|}-{\bf q}_{\|}|,\sigma^\prime})}
{\hbar\omega+E^{\rm(e)}_{|{\bf k}_{\|}-{\bf q}_{\|}|,\sigma^\prime}
-E^{\rm(e)}_{k_{\|},\sigma}+i0^+ }
\nonumber\\
&\times&\left[1+\sigma\sigma^\prime\,
\frac{k_{\|}-q_{\|}\cos(\theta_{\bf kq})}{|{\bf k}_{\|}-{\bf q}_{\|}|}
\right]\ ,
\label{eps}
\end{eqnarray}
$f_0$ is the Fermi function for electrons in a thermal-equilibrium state and $\theta_{\bf kq}$ is the angle between the wave
vectors ${\bf k}_{\|}$ and ${\bf q}_{\|}$.
It is convenient to make double transformation $\bf{k}_{\|} \rightarrow \bf{k}_{\|}+\bf{q}_{\|}$, $\bf{q}_{\|} \rightarrow -\bf{q}_{\|}$
in the second term of Eq.\,(\ref{eps}).
In the calculation
for image potentials, we only require the static polarization
function $\omega \rightarrow 0$.
For zero temperature regime it leads to:
\begin{eqnarray}
\label{eps11}
\alpha_{\rm L}(q_{\|},\,\omega)=\frac{e^2}{2\pi\hbar^2}\lim\limits_{\delta\rightarrow 0^+}\sum\limits_{\eta=\pm 1}\sum\limits_{\lambda=\pm 1}
\int\limits_{0}^{\infty} dk_{\|}\,g_{k_{\|}}\int\limits_{0}^{2\pi} d\theta_{\bf{kq}}\\
\nonumber
\times
\left[\frac{1+
\left({k_{\|}-q_{\|}\cos\theta_{\bf kq}}\right)/|{\bf k}_{\|}-{\bf q}_{\|}|}{E^{\rm(e)}_{k_{\|},\eta}-E^{\rm(e)}_{|{\bf k}_{\|}-{\bf q}_{\|}|,\eta}
+i\lambda\delta}
+\frac{1-
\left({k_{\|}-q_{\|}\cos\theta_{\bf kq}}\right)/|{\bf k}_{\|}-{\bf q}_{\|}|}{E^{\rm(e)}_{k_{\|},\eta}-E^{\rm(e)}_{|{\bf k}_{\|}-{\bf q}_{\|}|,-\eta}
+i\lambda \delta}\right]
\end{eqnarray}
Here, $g_{k_{\|}}$ are the cut-off functions for the intra- and inter-band excitations.
It is convenient to take it in the form of Gaussian:

\begin{eqnarray}
g_{k_{\|}}=\frac{1}{\sqrt{2\pi\sigma^2 (k_F+k_{\rm R})^2}}\,\exp\left[{-\frac{k_{\|}^2}{2 \sigma^2 (k_F+k_{\rm R})^2}}\right]
\label{cut-off}
\end{eqnarray}
with $k_F=\sqrt{2m^\ast E_F+k^2_{\rm R}}$ and $k_{\rm R}$ being defined in the next section.
Numerical simulations of the next section show that the polarization does not change perceptibly for the adjustable parameter $\sigma$ to be in the range $0.5 \leq \sigma \leq 0.85$.

\section{Numerical Results and Discussion}
\label{sec4}

In this section, we first introduce unitless quantities:

\begin{eqnarray}
k_{\rm R} &=& \frac{\Delta_{\rm R} m^\ast/\hbar^2}{\sqrt{2 \pi n_{e} +
(\Delta_{\rm R} m^\ast/\hbar^2)^2}}
\nonumber\\
q&=&\frac{q_{\|}}{2 \sqrt{2 \pi n_{e}
+(\Delta_{\rm R} m^\ast/\hbar^2)^2}}
\nonumber\\
r_s &= & \frac{m^\ast e^2}{\epsilon_0\,
\hbar^2 \sqrt{\pi n_{e}}}
\end{eqnarray}
where $m^\ast$ is the electron effective mass, $n_{e}$
is the surface electron density, $\Delta_{\rm R}$ is the Rashba
parameter, $r_s$ is the Wigner-Seitz radius which is the ratio
of the average inter-electron Coulomb interaction to the
Fermi energy $E_F\equiv\hbar^2k_F^2/(2m^\ast)=\hbar^2 \pi n_{e}/m^\ast$.
In terms of these  variables, we can rewrite
$\alpha_{\rm L}(q_{\|},\,0)/ 4 \pi \epsilon_0\,q_{\|}$
appearing in Eqs.\,(\ref{eels31}) and (\ref{eels31c.1}) in its unitless form:
$(r_s/q)\,(2 \pi/m^\ast)\,\Pi^0(q,\,0)$.
Here, the static polarization function obtained from
Eq.\,(\ref{eps}) with $\omega=0$ is given by the
closed-form analytic expressions.
Following the same procedure as in Ref.\,[\onlinecite{exact2}]
and setting the cut-off functions $g_k = \theta(k-\eta k_R)$ for 2DEG with SO we obtain:

\begin{eqnarray}
\label{EQ1}
&-&  \frac{2\pi}{m^\ast}\,\Pi^0(q,\,0)=
2\,\theta\left({1-k_{\rm R}-q}\right)+\theta\left(k_{\rm R}-\vert{q-1}\vert\right)\,
\left[1+\frac{\pi}{2}\,\sin(\psi)\right]
\nonumber\\
&-& 2\,\theta\left(q-1\right)\,\cosh^{-1}(q)\,\cos(\psi)+\sum\limits_{\eta=\pm1}\,
\theta\left(q-1-\eta k_{\rm R}\right)
\nonumber\\
&\times&
\left\{1+\eta\psi_\eta\,\sin(\psi)
-\cos(\psi_\eta)
-2\,\cos(\psi)\,
\ln\left[
\frac{1+q\,\sin(\psi_\eta - \eta\psi)}{2\sqrt{2q}\,\cos(\psi/2)\,\cos(\psi_\eta/2)}\right]\right\}\ ,
\label{polar}
\end{eqnarray}
where we have used $\sin(\psi)=(k_{\rm R}/q)\,\theta(q-k_{\rm R})$ and
$\sin(\psi_\eta)=(1+ \eta k_{\rm R})/q)\,\theta(q-1- \eta k_{\rm R})$.
\medskip

For the TI the Fermi energy crosses the dispersion curves in odd number of points, unlike for 2DEG with SO where the number of crossing points is even.
Therefore there are no intraband excitations for $\eta=-1$.
This can be modeled by the cut-off function $ g_{k_{\|}}$ in Eq.\,(\ref{cut-off}).
\medskip

\begin{figure}[p]
\begin{center}
\includegraphics[width=0.8\columnwidth]{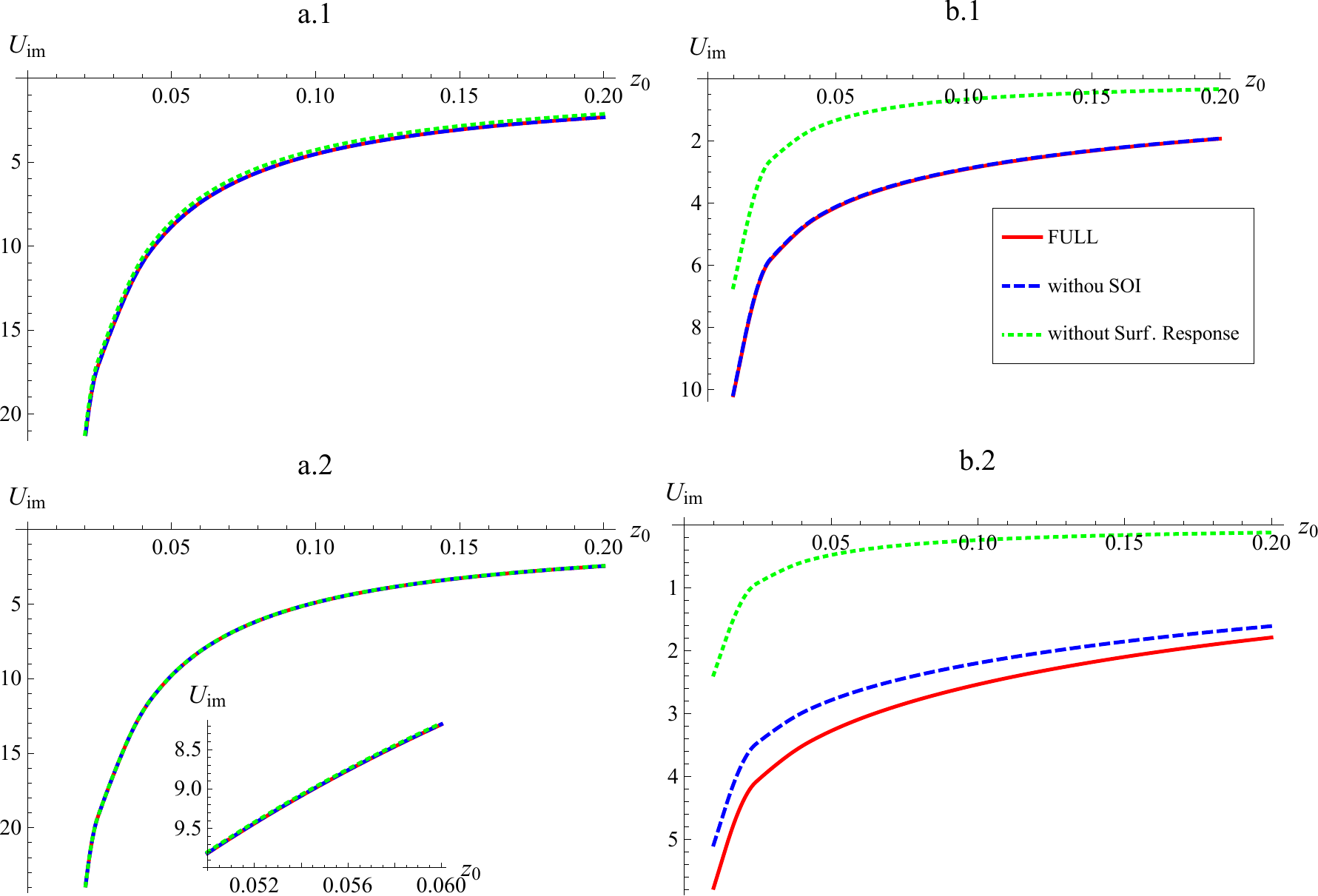}
\caption{(Color online) The image potentials $\mathcal{U}_{\rm im}$
[in units of $-e^2/(2\epsilon_0)\,\sqrt{2 \pi n_e
+(\Delta_{\rm R} m^\ast/\hbar^2)^2}$] as a function of
$z_0$ [in units of $1/\sqrt{2 \pi n_e
+(\Delta_{\rm R} m^\ast/\hbar^2)^2}$]
for the single-layer 2DEG and TI. Panels (a.1) and (b.1) correspond
to the 2DEG with $\epsilon_b = 13.1$ and $\epsilon_b=2.1$, respectively.
Panels (a.2) and (b.2) correspond to TI with $\epsilon_b = 100.0$ and $\epsilon_b=2.1$. The dotted curve shows the results for the  bare
image potential $V_0(z_0)=-[(\epsilon_b-1)/(\epsilon_b+1)]\,(1/4z_0)$
from Eq.\,(\ref{EQ:IM1}) for comparison. The solid and dashed
curves display the results for the exact image potential
${\cal U}_{\rm im}(z_0)$ constructed with the use of
Eq.\,(\ref{U-image}).}
\label{FIG:1}
\end{center}
\end{figure}

\begin{figure}[p]
\begin{center}
\includegraphics[width=0.8\columnwidth]{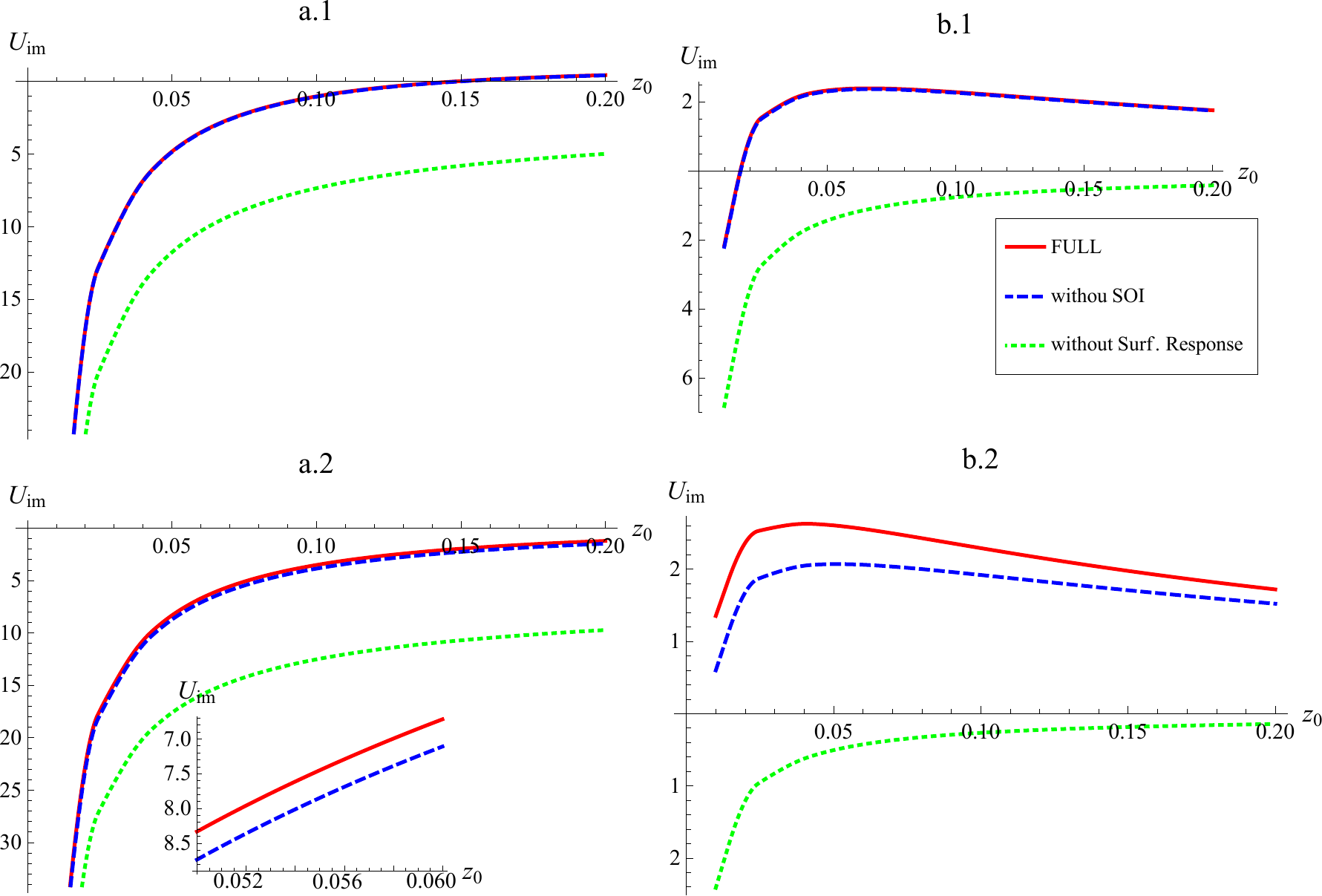}
\caption{(Color online)  The image potentials
$\mathcal{U}_{\rm im}$  [in units of $-e^2/(2\epsilon_0)\,
\sqrt{2 \pi n_e +(\Delta_{\rm R} m^\ast/\hbar^2)^2}$]
 as a function of $z_0$ [in units of $1/\sqrt{2 \pi n_e
+(\Delta_{\rm R} m^\ast/\hbar^2)^2}$]
for the double-layer 2DEG and TI. Panels (a.1) and (b.1) correspond
to the 2DEG with $\epsilon_b = 13.1$ and $\epsilon_b=2.1$, respectively.
Panels (a.2) and (b.2) correspond to TI with $\epsilon_b = 100$ and
$\epsilon_b=2.1$. The dotted curve exhibits the results for the bare
image potential $V_0(z_0)=-[(\epsilon_b-1)/(\epsilon_b+1)]\,(1/4z_0)$
from Eq.\,(\ref{EQ:IM1}) for comparisons. The solid and dashed
curves present the results for the exact potential
${\cal U}_{\rm im}(z_0)$ calculated from Eq.\,(\ref{U-image}).}
\label{FIG:2}
\end{center}
\end{figure}

The necessary condition for showing SOI effect on image potential is a large Rashba parameter $\Delta_{\rm R}$ and a strong effect from
a density-density response function, i.e., $\alpha_{\rm L}(q_{\|},\,0)/(4\pi\epsilon_0\,q_{\|})$ becomes comparable to $\epsilon_b$.
In Fig.\,\ref{FIG:1}, we plotted the image potential as a function
of $z_0$ for the single-layer 2DEG and TI. In addition, in
Fig.\,\ref{FIG:2}, we present results for 2DEG and TI
with a double-layer structure and a layer separation $d=100$\,\AA.
We have used Eq.\,(\ref{EQ1}) in the calculation of required $\alpha_{\rm L}(q_{\|},\,0)$ in the surface-response function as well as the following
numerical parameters.
The electron density is $n_e=1\times 10^{11}$\,cm$^{-2}$.
We chose the background dielectric constant appropriate
for GaAs/AlGaAs  as $\epsilon_b=13.1$ for the 2DEG,
and $100$ for the TI (Be$_2$Se$_3$). The remaining parameters
used in our calculations are
as follows. For the 2DEG: $k_{\rm R}/k_F=0.1747$  and the
Wigner-Seitz cell  parameter  $r_s=20.1$. For the TI:
$k_{\rm R}/k_F=0.99355$ and $r_s=0.771239$.
\medskip

For these chosen values, our calculations have discovered that
there is no difference for the image potential with/without
the Rashba SOI for either the single-layer or the double-layer TI and 2DEG
(with $d=100$\,\AA)
when the ratio of the background dielectric constants for
the materials on either side of the layers is large (i.e., $\epsilon_b=13.1$ for the 2DEG and $\epsilon_b=100$ for the TI) and the
point charge is placed in the region with a smaller dielectric constant.
It is clear from Eq.\,(\ref{eels31c.1}) that the surface-response function for the single-layer case (with $d\rightarrow\infty$) approaches unity as $\epsilon_b\gg 1$ and
$|\alpha_{\rm L}(q_{\|},\,0)|/(4\pi\epsilon_0\,q_{\|})\ll \epsilon_b$.
As a result, no effect from SOI, which is included in the density-density response function, should be seen.
For the double-layer case, the surface-response function in Eq.\,(\ref{eels31}) does deviate from unity as long as $q_{\|}d\ll 1$ is met.
However, $|\alpha_{\rm L}(q_{\|},\,0)|/(4\pi\epsilon_0\,q_{\|})\ll \epsilon_b$ still holds, implying no visible SOI effect.
In this case, the long-range part of the inter-layer Coulomb interaction dominates the density-density response function, leading to a spin-independent contribution to the image potential.
When $\epsilon_b=2.1$ is assumed, we find from two panels (b.1) of Figs.\,\ref{FIG:1} and \ref{FIG:2} for 2DEG that the effect of surface-response function is greatly enhanced
although the SOI effect is still invisible due to small Rashba parameter $\Delta_{\rm R}$ involved for the 2DEG. However, the
SOI effect for TI becomes much more significant as can be seen from two panels (b.2) of Figs.\,\ref{FIG:1} and \ref{FIG:2} due to a much larger Rashba-like  parameter involved for the TI.
Consequently, the effect of SOI may be manipulated by adjusting
the dielectric environment. For example, for the TI with a
Rashba-like term to be comparable with the quadratic term in
the energy dispersion due to a large value for $m^\ast_{TI}$,  Fig.\,\ref{FIG:2}
demonstrates a spin-dependent sign switching with decreasing $\epsilon_b$ for the double-layer TI system.
\medskip

\begin{figure}
\centering
\includegraphics[width=0.8\columnwidth]{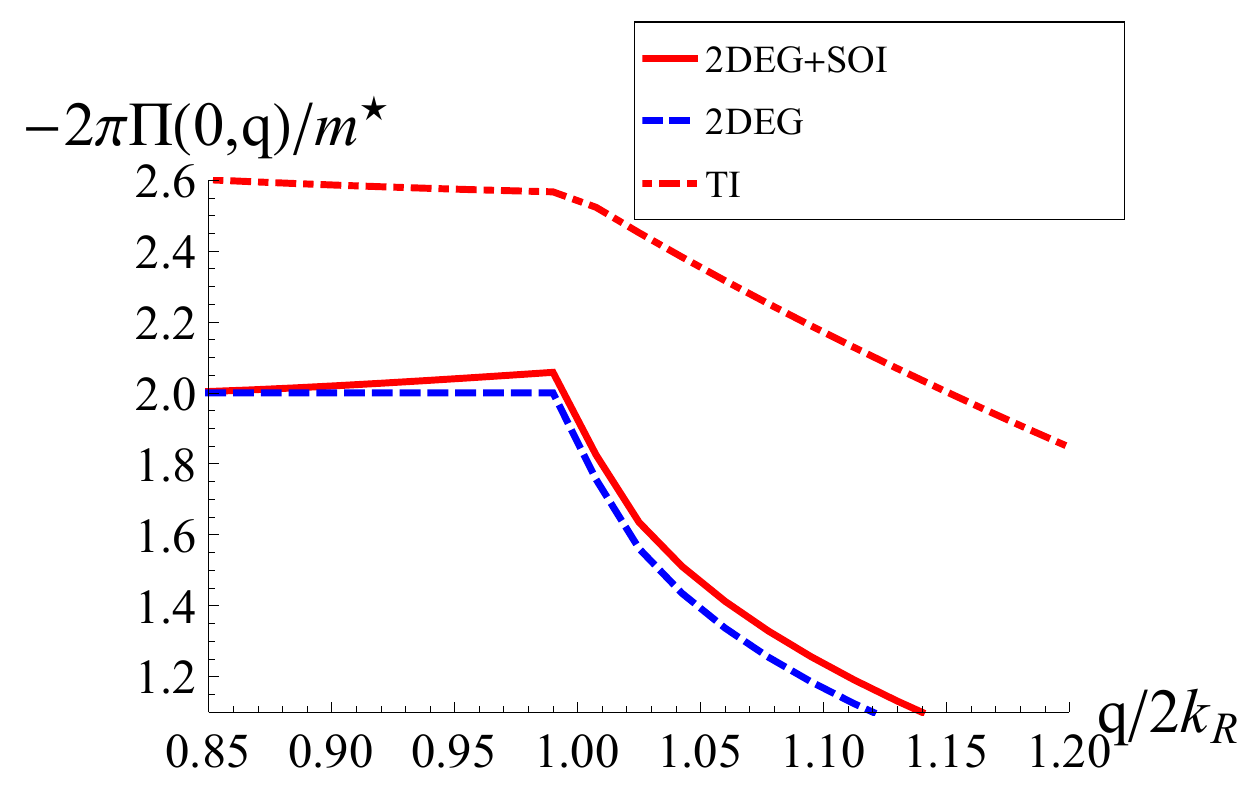}
\caption{\label{FIG:3} The dimensionless static polarization function $-2\pi\,\Pi(q,\,0))/m^\ast$ calculated from Eq.\,(\ref{polar}), as a
function of the scaled in-plane wave number $q/(2k_{\rm R})$. The plots
are for the 2DEG with/without spin-orbit interaction as well as for the
topological insulator with SOI. The parameters
used in the calculations are given in the text.}
\end{figure}

From the single-layer result presented in Fig.\,\ref{FIG:1} we find that the effect of surface response function becomes significant for small values of $\epsilon_b$. In addition, the SOI effect is more important for
TIs than for 2DEGs. For the double-layer result displayed in Fig.\,\ref{FIG:2}, we further observe that the inter-layer Coulomb interaction can enhance the the effect of surface response function and change the sign of
image potential from negative to positive. Again, the SOI effect is seen much more significant in TIs, where
the SOI effect is contained within the static density-density response function $\alpha_{\rm L}(q_{\|},\,0)=4\pi e^2\,\Pi^0(q_{\|},\,0)$ and is independent of the layer separation.
The higher the $\epsilon_b$ value is, the larger the asymmetry of the dielectric environment will be. Physically, the low values of $\epsilon_b$ implies a robustness of the system to an induced depolarization field by the
Coulomb field from an external point charge. This mechanism is reflected in the calculated surface response function. For double-layer 2DEG/TI systems, the capability of the system to screen the depolarization field
produced by the external point charge is further enhanced due to the existence of additional inter-layer Coulomb coupling. In order to see the role
of static density-density response function, we compare in Fig.\,\ref{FIG:3} the required static polarization function
$-2\pi\,\Pi^0(q_{\|},\,0)/m^\ast$ determined from Eq.\,(\ref{polar}) as a function of $q/(2k_{\rm R})$
for the 2DEG in the presence and absence of the linear Rashba term.
For comparison,  we also plotted $\Pi^0(q_{\|},\,0)$ for the TI with a built-in Rashba-like linear term.
The polarization function assumes a universal form for single
layer 2DEG and TI. The SOI effect on the polarization function
of the 2DEG is only enlarging it slightly. This minor enhancement becomes somewhat stronger
as $q$ is increased. Such an enhancement is also accompanied by an upward shift in the plasmon frequency in the long wavelength limit
for a 2DEG when SO coupling is included\,\cite{gg}.
For TI, on the other hand, $\Pi^0(q_{\|},\,0)$ becomes much stronger due to
a large Rashba-like  parameter. Furthermore,
the difference in the static polarization function becomes pronounced
for the image potential when the  dielectric environment is
adjusted as we described above.

\section{Concluding Remarks}
\label{sec5}

In this paper, we considered a point charge placed above a 2D conducting
layer at the interface between two dielectric media with  relative permittivity
$\epsilon_b$.  We have found from our study that the role played by the
SOI on the image potential may be modified by varying $\epsilon_b$. It
has been shown that the SOI makes a nontrivial difference  when the relative permittivity  between the two dielectric media is reduced, thereby
effectively separating the spin-`$\uparrow$' (+) and spin-`
$\downarrow$' ($-$) electrons  into different IQH states. Electron
spin is conserved in this case, but there is still a spin current.
The applied electric field due to the external point charge induces
oppositely directed Hall currents associated with `+' and `$-$'-spins,
respectively. Although the net charge current is zero,  the net spin
current is finite. The whole system becomes conducting due to
the metallic nature of the sample edges. This net spin current, however,
does not significantly affect the image potential of 2DEGs in our model.
In the work of Kane and Mele\,\cite{Kane-Mele}, it was shown
that, in real solids when all spins mixed, there is no net spin current.
It was further found that there exists a new topological invariant in
time-reversal-invariant systems of fermions. As a matter of fact,
a topological phase is insulating but always has metallic edges/surfaces
when the sample is put next to vacuum or an ordinary phase.  These
considerations motivated us to explore the 2DEG with SOI as well as
the 2D topological insulator using our formalism. Interestingly, for
the TI the role played by the helical linear energy dispersion on the
image potential is dramatically enhanced, in  comparison with 2DEG.
This occurs when  the relative permittivity of adjoined media is reduced.

\newpage
\begin{center}
{\bf Acknowledgments}
\end{center}

This research was supported by contract \# FA 9453-11-01-0263 of AFRL.
DH would like to thank the Air Force Office of Scientific Research (AFOSR) for its support.

\end{document}